\documentclass[aps,pra,reprint,showpacs]{revtex4-1}

\usepackage{amsmath}
\usepackage{amssymb}
\usepackage{graphicx}
\usepackage{hyperref}

\newcommand{\ket}[1]{|{#1}\rangle}

\newcommand{\R}[1]{{\textrm{#1}}}

\newcommand{\cT}{{\mathcal T}}

\begin{document}

\title{EIT-control of single-atom motion in an optical cavity}

\author{Tobias~Kampschulte} \email{kampschulte@uni-bonn.de}
\author{Wolfgang Alt}
\author{Sebastian Manz}
\author{Miguel Martinez-Dorantes}
\author{Ren\'{e} Reimann}
\author{Seokchan Yoon}
\author{Dieter Meschede}
\affiliation{Institut f{\"u}r Angewandte Physik, Universit{\"a}t
Bonn, Wegelerstra{\ss}e~8, D-53115~Bonn, Germany}
\author{Marc Bienert}
\author{Giovanna Morigi} \affiliation{Theoretische Physik, Universit\"{a}t des
Saarlandes, D-66123 Saarbr\"{u}cken, Germany.}
\date{\today}

\begin{abstract}
We demonstrate cooling of the motion of a single neutral atom confined by
a dipole trap inside a high-finesse optical resonator. Cooling of
the vibrational motion results from EIT-like interference in an
atomic $\Lambda$-type configuration, where one transition is
strongly coupled to the cavity mode and the other is driven by an
external control laser. Good qualitative agreement with the
theoretical predictions is found for the explored parameter
ranges. Further we demonstrate EIT-cooling of atoms in the dipole trap in free space, reaching the ground state of axial motion. By means of a direct comparison with the cooling inside the resonator, the role of the cavity becomes evident by an additional cooling resonance. These results pave the way towards a controlled interaction between atomic, photonic and mechanical degrees of freedom.
\end{abstract}


\maketitle

\section{Introduction}
Single emitters strongly coupled to optical resonators form a
promising basis for the realization of quantum networks
\cite{Cir97,Kim08}.  Experimental implementations include trapped
ions \cite{Stute2012}, atoms \cite{Mil05,Wil07,Ritter12},
artificial superconducting qubits \cite{Wallraff2004}, and
opto-mechanical devices \cite{Ver12}.  In these platforms, the
strong coupling between photons and emitters is utilized to
coherently transfer quantum information between stationary qubits
at the nodes and flying qubits acting as interconnects. A
prerequisite for high-fidelity operations is the control of the
coupling with the cavity mode, which implies spatial localization
of the emitter at the sub-wavelength scale. This requirement is
easy to fulfill for artificial atoms bound to a substrate. In ion
traps the steep confinement puts relatively low requirements on
cooling~\cite{Guthoerlein2001,Mundt2002,Keller2004}, while
realizations based on neutral atoms in dipole traps often require
to prepare the motion close to the vibrational ground state. The
latter thus call for efficient cooling techniques, which are
robust and sufficiently fast to enable viable quantum
technological implementations.

Ground-state cooling of trapped atoms usually makes use of a narrow
resonance, which allows one to selectively address transitions
where scattering induces the loss of a quantum of
vibration~\cite{eschner2003}. Such resonances can be realized by
choosing a suitable dipolar transition, as in sideband
cooling. In Raman sideband cooling or EIT cooling, narrow transitions are obtained by coherent two-photon coupling of two stable states~\cite{Mor00,Roo00,Mor03,Sch01,Lin13}. One possibility to cool atoms inside a cavity is to perform Raman sideband cooling there,
as shown in Ref.~\cite{Boo06,Rei13}. The corresponding experimental
effort, however, notably increases with the number of degrees of
freedom to cool. An alternative is to exploit the strong coupling
at the single-photon level. In this case, the relevant narrow
resonance is determined by the finite cavity
lifetime~\cite{Vuletic2001}, thus implementing a form of sideband
cooling even on dipolar transitions whose radiative linewidth in
free space is broader than the trap frequency~\cite{Lei09}. In
fact, the cavity plays a similar role of an additional resonance with linewidth
$2\kappa$~\cite{Zip05,Bienert2012b}. Sub-Doppler or even sub-recoil cooling \cite{Wol12}
can then be realized if the cavity linewidth $2\kappa$ is smaller than the
trap frequency $\omega$. For typical dipole traps this condition implies
rather ‘closed’ cavities (with a small $\kappa$), where coupling photons
in and out is relatively slow, and which are thus less suitable for fast and efficient interfaces. 

In this report we experimentally characterize cooling of the vibrational motion of a
trapped atom by driving a high-finesse cavity coupled to a
three-level $\Lambda$ transition. This work was triggered by recent experimental studies of single photon EIT~\cite{Muc10,Kam10},
where long trap lifetimes times have been observed~\cite{Kam10},
indicating robust cooling in this system, and on recent
theoretical studies~\cite{Bienert2012} of cooling in this
experimental setup. Here, we demonstrate cavity-assisted
EIT cooling of an atom in a dipole trap. The scheme relies on the frequency selectivity of narrow dark resonances, thus the final temperature is not limited by the resolved sideband condition ($2\kappa\ll\omega$). Moreover, we exploit the
possibility to switch between EIT-like cooling and cavity cooling
by simply changing the detuning of one laser to compare the
cooling efficiency in the different regimes, and we show that the
interplay of Raman and cavity resonances gives rise to novel and
robust cooling regimes. We show that theoretical predictions and
experimental results are in remarkable agreement.


\begin{figure*}
  {(a)}\hspace{35mm}{(b)}\hspace{37mm}{(c)}\hspace{40mm}{(d)}\hspace{40mm}\mbox{}\\
  \vspace{-3mm}
  \includegraphics[width=0.19\textwidth]{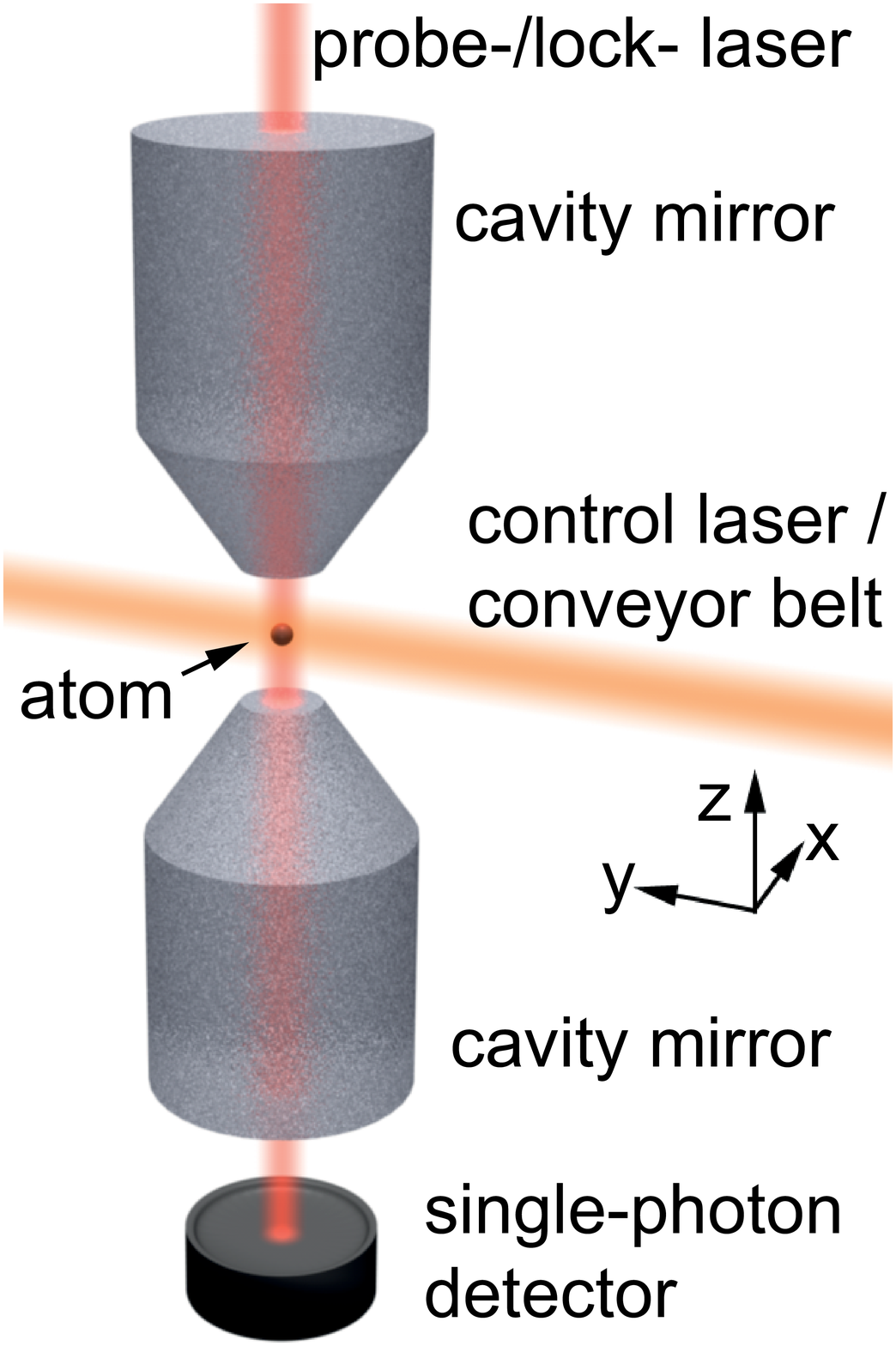}
  \hspace{5mm}
  \includegraphics[width=0.75\textwidth]{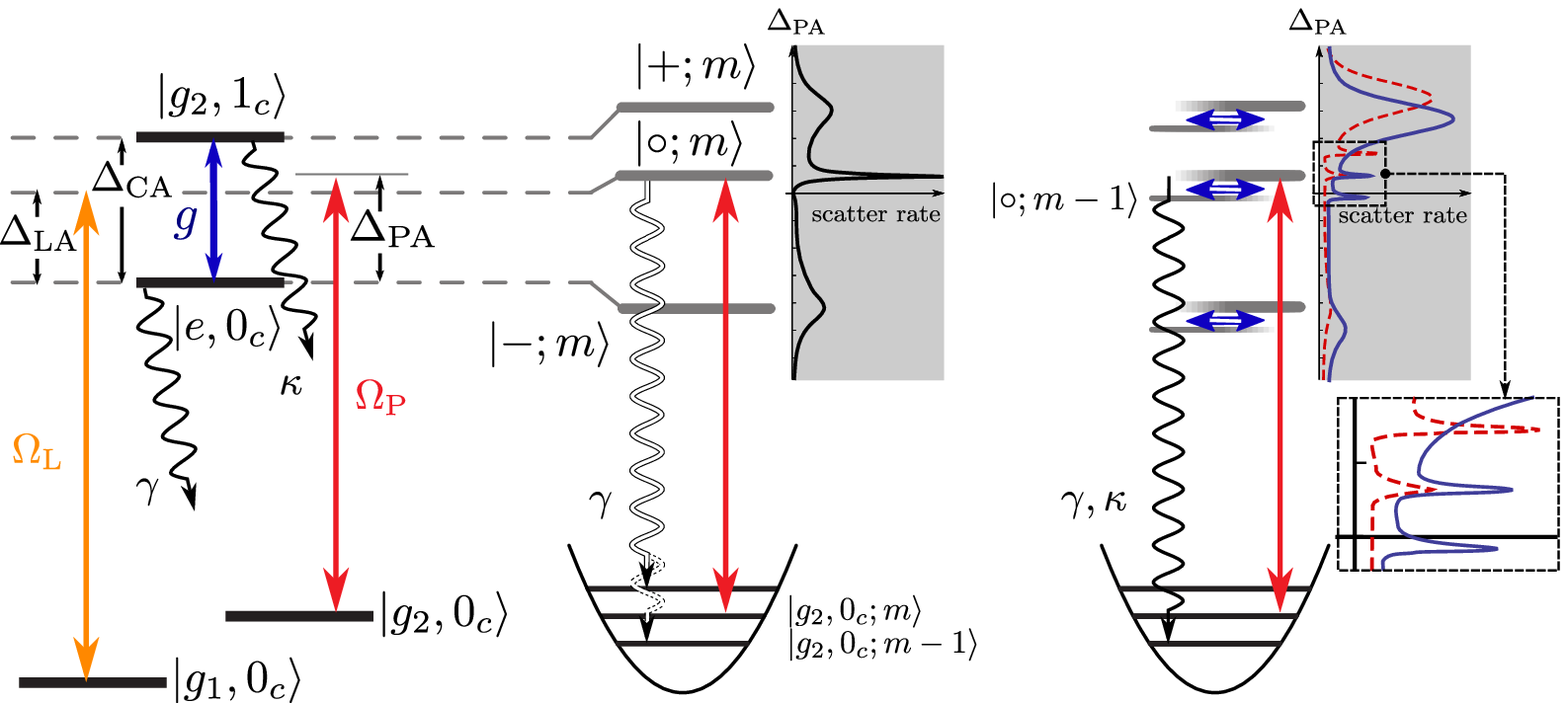}
\caption{(Color
  online) (a) Schematic experimental setup. A single cesium atom is
trapped inside a high-finesse optical cavity and illuminated by a
control laser along the dipole trap axis. A probe laser is coupled
into the cavity mode and its transmission is detected by a single
photon detector. (b) Relevant bare energy levels (black) of atom
and cavity, Rabi frequencies and detunings of lasers and cavity
mode. (c,d) Dressed states (grey) and scattering processes leading
to cooling. The vibrational excitations are shown explicitly. The
straight (wiggled) arrows represent the coherent (incoherent)
transitions which lead to a change of the energy by one
vibrational quantum $\hbar\omega$. Hollow arrows identify the
process which mediates the mechanical action. (c) Cooling
(heating) transition due to the recoil of fluorescence photons.
Inset: excitation spectrum, determining the corresponding rate, as
a function of the probe frequency. This process is diffusive and
can be suppressed using EIT. (d) Cooling transition due to the
mechanical effect of the cavity field (blue arrows). The inset
displays the corresponding transition rate (blue curve). The red
dashed curve corresponds to heating. The double Fano structure is
a consequence of the strong atom-cavity
coupling~\cite{Bienert2012}. Diffusion and cooling/heating add up
to give the rates $A_{\pm}^{(j)}$.
 \label{level_scheme}}
\end{figure*}

\section{Experimental Setup}

In our experiment, Fig.~\ref{level_scheme}(a), single laser-cooled
neutral cesium atoms are loaded from a magneto-optical trap (MOT, $852\,$nm)
into a far-red-detuned standing-wave dipole trap (DT, $1030\,$nm)
with an axial trap frequency of $\omega_\text{DT}/2\pi=0.3\,$MHz
along the $y$-direction and a radial trap frequency of about $1\,$kHz~\footnote{The weak radial confinement allows for a slow variation of the atom-cavity coupling strength, see ref.~\cite{Khu08}}. After verifying that exactly one atom has
been loaded and determining its position, it is subsequently
transported, using the DT as a conveyor belt, into the mode of the high-finesse optical cavity, for details see~\cite{Khu08}. The cavity resonance
frequency is stabilized using a blue-detuned lock-laser
($845\,$nm), which forms at the same time a standing-wave dipole
trap along the cavity axis and thus confines the atomic motion
along the $z$-direction with an axial trap frequency of about
$\omega/2\pi= 0.2\,$MHz.
The relevant electronic levels are the long-lived hyperfine ground states
$\ket{g_{1}}\equiv\ket{6^2\text{S}_{1/2},F=3}$ and
$\ket{g_{2}}\equiv\ket{6^2\text{S}_{1/2},F=4}$, which couple to
the common excited  state
$\ket{e}\equiv\ket{6^2\text{P}_{3/2},F=4}$ via an external control
laser (propagating along the $y$-direction) and the cavity mode
(along $z$), respectively.
The control laser drives the transition
$\ket{g_1}\leftrightarrow\ket{e}$ with effective strength
$\Omega_\text{L}=2\pi\times2.8\,$MHz and variable detuning
$\Delta_\text{LA}$, see Fig.~\ref{level_scheme}(b). The cavity
field has vacuum Rabi frequency $g=2\pi\times3.6\,$MHz at its
antinodes and a detuning of
$\Delta_\text{CA}=2\pi\times(16\pm3)\,$MHz from the atomic
transition $\ket{g_2}\leftrightarrow\ket{e}$. The cavity is weakly
driven by a probe laser with variable detuning $\Delta_\text{PA}$
from the transition $\ket{g_2}\leftrightarrow\ket{e}$. We denote
by $\delta_\text{PC}\equiv\Delta_\text{PA}-\Delta_\text{CA}$ its
detuning from the empty cavity resonance. At $\delta_\text{PC}=0$
the probe-laser strength $\Omega_\text{P}=2\pi\times0.23\,$MHz
corresponds to a mean intracavity photon number of $0.08$. Since $g$ is larger than the atomic dipole and cavity
field decay rates $(\gamma,\kappa)=2\pi\times(2.6,0.40)\,$MHz,
the system is in the strong
coupling regime, and for our parameters, one atom in state $|g_2\rangle$ causes the
cavity transmission to drop by about $50\%$ compared to the on-resonance transmission.

\section{Theoretical description} 

The weak-excitation regime allows us to assume that the relevant
states of the cavity field are $\ket{n_\text{c}}$ with
$n_\text{c}=0,1$ photons. The model for the atom is based on three
levels ignoring that each hyperfine state is degenerate, which
will be sufficient to qualitatively describe the experimental
results. Therefore, the basis of the atom-photon system consists
in the states
$\ket{g_{2},0_\text{c}},\ket{g_{1},0_\text{c}},\ket{e,0_\text{c}},\ket{g_{2},1_\text{c}}$,
which are coupled as depicted in Fig.~\ref{level_scheme}(b). State
$\ket{g_{2},0_\text{c}}$, in particular, weakly couples to state
$\ket{g_{2},1_\text{c}}$, which in turn strongly couples to the
other states. The excitation spectrum with respect to the probe
laser shows that the level structure is conveniently described in
terms of $\ket{g_{2},0_\text{c}},$ and the dressed states
$\ket{\pm}$ and $\ket{\circ}$, {\it i.e.} the lowest-energy excited eigenstates of the control-laser driven atom-cavity system neglecting the motion~\cite{Bienert2012}, see Fig.~\ref{level_scheme}(c). 
These dressed states have eigenfrequencies $\omega_j$ with $j=\pm,\circ$ and are superpositions of the basis states $\ket{g_1,0_{\rm c}}$, $\ket{g_2,1_{\rm c}}$ and $\ket{e,0_{\rm c}}$. Since the latter two states are radiatively unstable, a finite linewidth $\gamma_j$ can be associated with each dressed state. 

A Fano-like profile is observed (inset of
Fig.~\ref{level_scheme}(c)) when the detuning
$\delta_\text{PL}\equiv\Delta_\text{PA}-\Delta_\text{LA}$
vanishes, namely, when $|g_2,0_{\rm c}\rangle$ and $|g_1,0_{\rm
c}\rangle$ are resonantly coupled in a three-stage process: In
this case the system exhibits a dark state where $\ket{e,0_{\rm
c}}$ is not excited. Our four-level system is thus similar to the
one discussed in~\cite{Champenois06}, whereby here one excitation
is purely photonic~\cite{Bienert2012}.

An example of scattering processes leading to cooling and heating
along the cavity axis is sketched in Fig.~\ref{level_scheme}(c)
and (d). Setting the probe frequency close to a narrow resonance induces 
processes where a vibrational quantum can be lost by either a
diffusive process associated with the mechanical effect of the
flourescence photon, (c), or by the mechanical force of the cavity
field, (d). The two kinds of processes add up  giving the total
cooling (heating) rate. The existence of dark states leads to Fano
resonances in the scattering rates and can be exploited to
suppress diffusion and heating processes. 

The cooling dynamics is theoretically modeled approximating the dipole potential by a harmonic oscillator and using the rate-equation description of Ref.~\cite{Bienert2012} in a one-dimensional treatment~\footnote{Within the Lamb-Dicke limit, no coupling between different spatial directions is present. Hence, the cooling along the traps in $z$- and $y$- direction can be described independently}. The model is valid for small photon numbers $\langle n\rangle\ll 1$ inside the cavity and for weak mechanical coupling, characterized by a small Lamb-Dicke parameter $\eta = [\omega_{\rm rec}/\omega]^{1/2}\ll 1$, where $\omega_{\rm rec}$ is the atomic recoil frequency and $\omega$ is the trap frequency. The theory delivers the heating and cooling rates $A_\pm$
connected with photon scattering along the blue and red sideband transitions $\ket{g_2,0_{\rm c},m}\to \ket{g_2,0_{\rm c},m\pm 1}$, respectively, along the axis of motion. For cooling along the cavity axis ($z$-axis) in the configuration of Fig.~\ref{level_scheme}(a), the rates can be written in the form
\begin{align}
 A^{(z)}_\pm = 2\gamma \mathcal{D}  + 2\gamma |\cT_\pm^{\gamma,{\rm C}}|^2 + 2\kappa |\cT_\pm^{\kappa,{\rm C}}|^2.
 \label{eq:Apm}
\end{align}
The first term in Eq.~\eqref{eq:Apm}, proportional to $\gamma$ and identical for heating and cooling,
describes the diffusion of the atomic motion due to mechanical
effects of spontaneous decay, and corresponds to the process sketched in Fig.~\ref{level_scheme}(c). The second (third) term stems from
processes where light is scattered by the atom (cavity), whereby
the mechanical action is due to the Jaynes-Cummings coupling, Fig. ~\ref{level_scheme}(d). The exact form of the quantities $\mathcal{D}$, $\cT_\pm^{\gamma,{\rm C}}$ and $\cT_\pm^{\kappa,{\rm C}}$, and the form of $A_\pm$ for cooling along other directions can be found in \cite{Bienert2012}.

From the quantities $A_\pm$ the cooling rate 
\begin{align}
 \Gamma = A_- - A_+
 \label{eq:Gamma}
\end{align}
and, if $\Gamma>0$, the mean occupation number
\begin{align}
 \langle m\rangle = \frac{A_+}{A_--A_+}.
\end{align}
in the stationary (thermal) state at the end of cooling can be calculated~\cite{Stenholm86}. 
The cooling rate $\Gamma$ gives the time scale at which the stationary state of motion is reached.

The general behavior of the
rates~\eqref{eq:Apm} is determined by the states $\ket{\pm}, \ket{\circ}$:
$A_-$ ($A_+$) is strongly enhanced,
whenever the probe laser is resonant with the red (blue) sideband
of one of the dressed states, {\it i.e.} for $\delta_\text{PC} = \omega_j
-\omega$ ($\delta_\text{PC} = \omega_j +\omega$). The linewidth $\gamma_j$ of the dressed states determine the width of these resonances and therewith the cooling dynamics when the probe is tuned to the red sideband. In particular, for $\gamma_j\ll\omega$, resolved sideband cooling at the corresponding dressed state resonance can be performed.

EIT-like cooling in the cavity can be observed when the resonance
 condition $\delta_{\rm PC} = \Delta_{\rm LA}-\Delta_{\rm CA}$ is fulfilled.  The appearance of a
dark state follows from quantum interference between the three
dressed states in
the excitation process of the cavity-atom system, and occurs when
the ground state $\ket{g_2,0_{\rm c}}$ is resonantly coupled by three
photons (probe, cavity, laser) with $\ket{g_1,0_{\rm c}}$. In this case,
the rates are
\begin{align}
A^{(z)}_\pm =& \frac{\Omega_{\rm P}^2/2}{\delta_{\rm PC}^2+\kappa^2}\eta^2\sin^2(k x_0) g^2\gamma\times\nonumber\\
&\frac{1+\mathcal C_\pm} {\gamma^2\left(1+\mathcal
C_\pm\right)^2 + \left(\frac{\Omega_\R{L}^2}{4\omega}-\omega\pm
\Delta_{\rm LA}+ \frac{\gamma}{\kappa}\mathcal C_\pm(\omega \mp
\delta_{\rm PC})\right)^2}\,, \label{eq:ApmEIT}
\end{align}
with ${\mathcal C_\pm} =
C \kappa^2/{(\kappa^2+(\delta_{\rm PC}\mp\omega)^2)}$, whereby $C=\tilde
g^2/(\kappa\gamma)$ is the single atom cooperativity. The effective coupling constant $\tilde g = g \cos k x_0$ is proportional to the vacuum Rabi-frequency $g$ at an antinode of the cavity's cosinusoidal mode function of wavenumber $k$ with $x_0$ measuring the distance of the trap center from the antinode. 
 For $\Delta_{\rm LA} = \Delta_{\rm CA}$, three photon
 resonance is found for $\delta_{\rm PC} = 0$, and the rates
 Eq.~\eqref{eq:ApmEIT}
 take on the form of EIT cooling in free space with
 modified Rabi-frequency $\Omega_{\rm L}^2\to\Omega_{\rm L}^2 +
4\omega^2[\gamma'-\gamma]/\kappa$ and a modified atomic linewidth
$\gamma'=\gamma [C \kappa^2/{(\kappa^2+\omega^2)}+1]$. Hence, due the cavity-boosted Rabi-frequency, lower temperatures compared to free space EIT cooling are possible. Similarly, for large $\delta_{\rm PC}$ (and around $\delta_\text{PL}=0$), one recovers normal EIT-cooling~\cite{Mor00} in a standing wave, where the width of the EIT resonance determines the cooling. For the parameters of the experiment, this sets the limit for lowest achievable temperatures. To fully enter a regime where quantum interference involving the cavity determines the final vibrational occupation number given by $C^{-1}$, $\kappa\ll\omega$ is required~\cite{Bienert2012}, a condition not being fulfilled here. Nevertheless, signatures of such a cooling scheme can be found around the range $|\delta_{\rm PC}|\lesssim\omega$, as discussed below.

\section{Measurements}

Experimentally, the cooling efficiency
is characterized by means of the survival probability
$P_\text{s}$, which is measured by first trapping a single atom
inside the cavity for a certain holding time $t$, then retrieving
it from the cavity using the conveyor belt and finally detecting
its presence by the MOT fluorescence. After many repetitions, the
measured $P_\text{s}$ is given by the ratio between final and
initial number of atoms. Radiative cooling and heating effects
become obvious when we compare $P_\text{s}$ with the value for
atoms in the absence of near-resonant light: In the presence of
the far-detuned trapping lasers only (DT and lock laser),
$P_\text{s}$ decays exponentially with a time constant of about
$120\,$ms, limited by parametric heating \cite{Geh98} due to
technical intensity fluctuations mainly of the intracavity lock
laser power.

\begin{figure}
\centering{\includegraphics[width=\columnwidth]{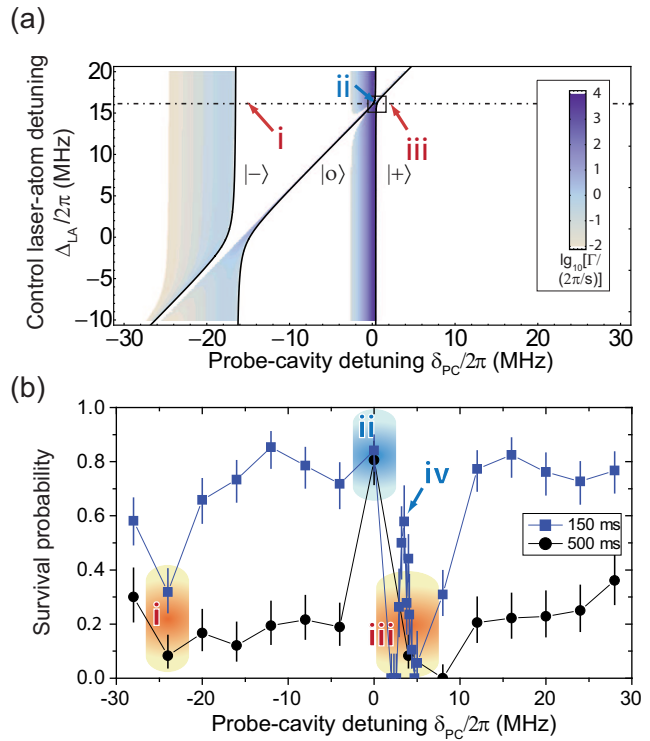}}\\
\caption{(Color
  online) (a) Cooling rate $\Gamma$, calculated for a fixed
cavity-atom detuning $\Delta_\text{CA}=2\pi\times16\,$MHz
according to \cite{Bienert2012}. The horizontal dashed-dotted line
corresponds to $\delta_\text{LC}=0$. The black solid lines
indicate the frequencies of the dressed states
$\ket{\pm},\ket{\circ}$. Away from the avoided crossing at
$\delta_\text{LC}=0$ the dressed states $\ket{\pm}$ are coherent
superposition of $\ket{g_2,1_\text{c}}$ and $\ket{e,0_\text{c}}$
which are coupled by the cavity field, whereas state $\ket{\circ}$
asymptotically coincides with $\ket{g_1,0_\text{c}}$. The small
box corresponds to the detail shown in Fig.~\ref{narrow_range}.
(b) Measurement of the single-atom survival probability as a
function of the probe-cavity detuning $\delta_\text{PC}$ for the
case $\delta_\text{LC}=0$ and for two different holding times $t$
inside the cavity. The parameter space corresponds to the
horizontal dashed-dotted line in (a).} \label{wide_range}
\end{figure}

We compare the survival probability with the theoretical
predictions for the cooling rate $\Gamma\equiv
A_-^{(z)}-A_+^{(z)}$, determining the time scale at which the
average number of vibrational excitation along the cavity axis
exponentially approaches the stationary value \cite{Stenholm86}.
Figure~\ref{wide_range}(a) displays a two-dimensional plot of
$\Gamma$, Eq.~\eqref{eq:Gamma}, using the rates from Eq.~\eqref{eq:Apm}. Here, the detunings $\delta_\text{PC}$
and $\Delta_\text{LA}$ are varied and $\Delta_\text{CA}$ is at the
fixed experimental value. Large cooling rates are predicted mostly
when the probe laser is tuned to the red side of a dressed-state
resonance (shaded areas).

Cooling and heating effects close to the dressed states
$|\pm\rangle$, $|\circ\rangle$ become experimentally visible when
scanning the probe-cavity detuning $\delta_\text{PC}$ over a wide
range while
$\delta_\text{LC}\equiv\Delta_\text{LA}-\Delta_\text{CA}=0$
remains fixed (dashed-dotted line). Figure~\ref{wide_range}(b)
displays the corresponding survival probabilities $P_\text{s}$ for
two different holding times. The heating regions are denoted by
\textsf{(i)} and \textsf{(iii)}. Strong losses in \textsf{(iii)}
occur mostly at $\delta_\text{PC}/2\pi\approx (2\ldots8)\,$MHz,
where the laser is tuned on the blue side of the dressed state
$|+\rangle$ and heats the motion. In \textsf{(i)} the detuning is
$\delta_\text{PC}/2\pi\approx -25\,$MHz and the theory predicts
that the motion is outside the Lamb-Dicke regime. The cooling regions are
denoted by \textsf{(ii)} and \textsf{(iv)}. In \textsf{(iv)}, the
detuning $\delta_\text{PC}/2\pi\approx 3.5\,$MHz corresponds to
the detuning of the orthogonally polarized cavity mode separated
by the birefringence splitting, which is not taken into account by
the theoretical model. The maximum value of $P_\text{s}$ occurs in
the narrow region \textsf{(ii)} around $\delta_\text{PC}=0$ and
$\delta_\text{PL}=0$, see also Ref.~\cite{Kam10}.

\begin{figure}
\centering{\includegraphics[width=\columnwidth]{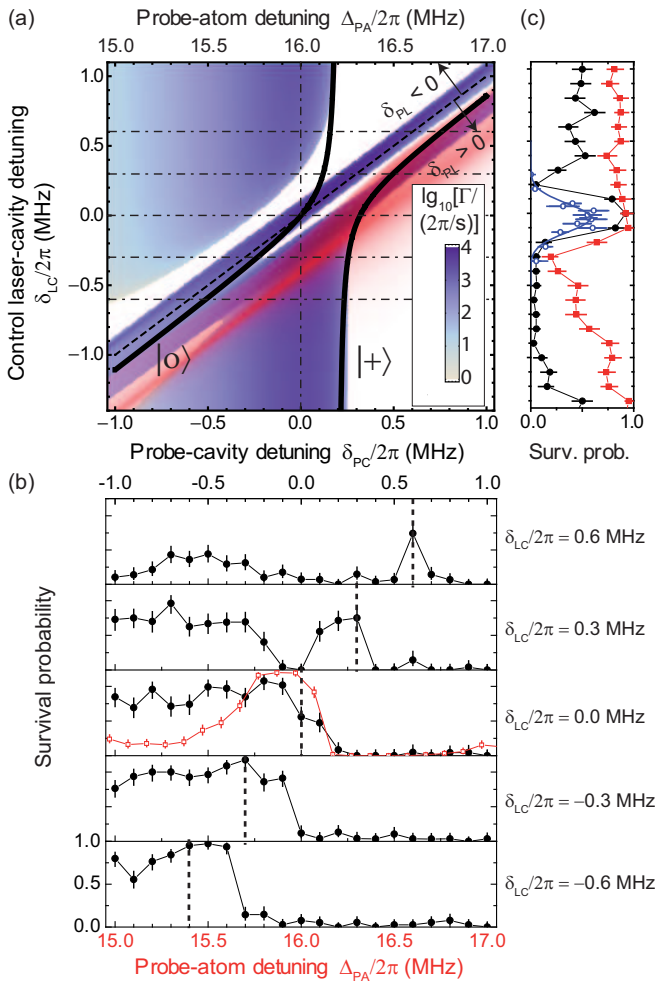}}\\
\caption{(Color
  online) (a) Calculated cooling rate along the cavity axis (blue
shading) and heating rate along the control laser axis (red
shading). Solid lines indicate the dressed-state energies of
$|\circ\rangle$ and $|+\rangle$ while the dashed diagonal line
corresponds to $\delta_\text{PL}=0$, where $A_\pm$ are given by Eq.~(\ref{eq:ApmEIT}). Horizontal and vertical
dashed-dotted lines correspond to the measurements in (b) and (c),
respectively. (b)~Survival probability of one atom as a function
of the probe-cavity detuning $\delta_\text{PC}$ for five different
values of $\delta_\text{LC}$ as indicated in (a) for a holding
time of $t=1\,$s  (black lines). Dashed lines indicate the
condition $\delta_{\rm PL}=0$. For comparison, the red line
indicates the survival probability as a function of the probe-atom
detuning $\Delta_\text{PA}$ for the case of EIT-cooling outside
the cavity. (c)~Survival probability of one atom at probe-cavity
resonance $\delta_\text{PC}=0$ as a function of $\delta_\text{LC}$
for three different holding times $t=0.02$, $0.3$ and $15\,$s
(red, black, blue).} \label{narrow_range}
\end{figure}

Figure \ref{narrow_range}(a) displays a zoom of the behavior of
$\Gamma$ in the boxed region of Fig. \ref{wide_range}(a): A rapid
change between cooling and heating regions is predicted on the
scale of the trapping frequencies. Lowest temperatures are
predicted along the diagonal blue stripe at largest $\Gamma$,
corresponding to $\delta_{\rm PL}=0$, with $\langle m_{\rm
st}\rangle\approx 0.07$ in the middle and down to $0.05$ at the
edges of the plot. Moreover, cooling is found in a second diagonal
stripe, where $\delta_\text{PL}\approx \omega$, and in
two regions above and below the stripes where
$\delta_\text{PC}\lesssim0.2\,\text{MHz}\times2\pi$.

We experimentally verify this behavior by scanning the
probe-cavity detuning $\delta_\text{PC}$ over $2\pi\times 2\,$MHz
(i.e., a few $\omega_\text{DT},\omega$) around
$\delta_\text{PC}=0$ for different values of $\delta_\text{LC}$.
The results are displayed in Fig.~\ref{narrow_range}(b): Cooling
is found around $\delta_\text{PL}=0$. It is also found when the
probe laser is resonant or red-detuned from the cavity frequency
($\delta_\text{PC}\lesssim0$) and, at the same time,
$\delta_\text{PL}<0$. Heating occurs for
$0\lesssim\delta_\text{PL}\lesssim1\,$MHz. This is also visible in
Fig.~\ref{narrow_range}(c), where the survival probability as a
function of  $\delta_\text{LC}$ is displayed \cite{Kam10}. These
observations can be understood if we consider that the lifetime of
atoms exposed to the lock laser potential is much shorter than the
holding time, hence the atoms only survive if they are cooled
along the cavity direction and not heated along the control laser
axis. The model predicts heating along this orthogonal axis for
$\delta_\text{PL}>0$, which is maximum for
$\delta_\text{PL}\approx \omega_\text{DT}$, see red shading in Fig.~\ref{narrow_range}(a). This can explain the
atom losses in this regime and why we see a single cooling peak at
$\delta_\text{PL}=0$ only, corresponding to the upper diagonal
stripe. 
This stripe corresponds to the EIT condition, where cooling is modified due to the presence of the cavity, and which is relevant at the level of a single photon already. Interference in the processes of scattering involving the cavity manifests itself by the alternating heating and cooling regions around the EIT stripe and the variation of the cooling rate along the stripe, which, for example, would allow to efficiently suppress heating transitions if the cavity sidebands were resolved.
We denote the mechanism by
cavity-assisted EIT (CEIT) cooling. The cooling at
$\delta_\text{PC}\lesssim0$, instead, is based on the excitation
of the red sideband of the cavity-like dressed state $\ket{+}$
\cite{Dom03,Zip05}. Here, in the limit of large
$|\delta_\text{PL}|$, the control laser acts like an incoherent
repumper that transfers atoms that decay into state $\ket{g_1}$
back to state $\ket{g_2}$ and the system can be treated as an
effective two-level system \cite{Kam10,Bienert2012b}.

\begin{figure}
\centering{\includegraphics[width=\columnwidth]{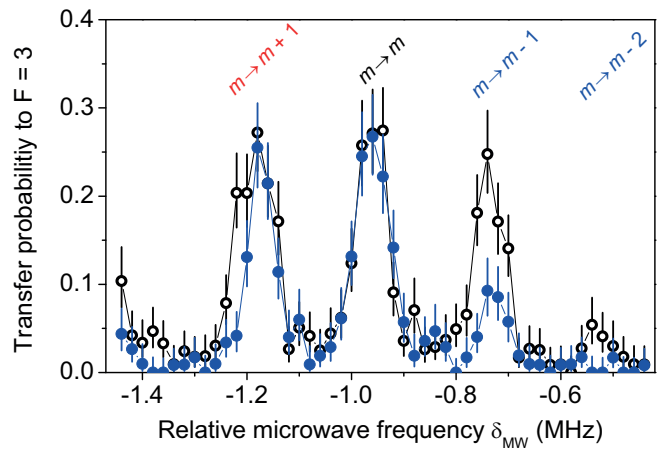}}\\
\caption{(Color
  online) Microwave sideband spectra of EIT-cooled (blue) and
molasses-cooled (black) atoms in the standing-wave dipole trap. For this, the cooled atoms were prepared in the Zeeman sublevel $\ket{F=4, m_F=-4}$ and transferred to $\ket{F=3, m_F=-3}$ using a microwave pulse with fixed power and duration. The transfer probability is determined by state-selective detection of atoms in $F=3$ and plotted as a function of the detuning $\delta_\text{MW}$ of the microwaves from the cesium clock transition frequency. The peak at $\delta_\text{MW}\approx-1$\,MHz corresponds to the carrier transition $m\rightarrow m$.
The sideband transitions $m\rightarrow m-1$ and $m\rightarrow m-2$ are strongly reduced in the
EIT case, indicating a high occupation of the vibrational ground state.} \label{MW_spectra}
\end{figure}

In order to clarify the role of the cavity in CEIT cooling, we
compare the observed dynamics with EIT cooling in free space. For
this purpose we measure the survival probability with small
ensembles of atoms trapped in the standing-wave dipole trap
outside the cavity (at the position of the MOT). Here, the probe
laser ($\Omega_\text{p}/2\pi=0.5\,$MHz) is directly shone on the
atoms, which are strongly confined only along the control-laser
axis. The holding time of $300$\,ms is much shorter than the trap
lifetime of several seconds. The atoms only survive within a
narrow region around $\delta_\text{PL}=0$ (EIT condition), see the
central plot of Fig.~\ref{narrow_range}(b).

To quantify the cooling performance of EIT-cooling in the dipole trap
and to compare it with standard molasses-cooling, we apply microwave sideband spectroscopy by making the 
lattice slightly state-dependent \cite{Foe09}. Fig.~\ref{MW_spectra}
shows spectra taken with EIT- and molasses-cooled atoms, respectively.
The significant reduction of the $m\rightarrow m-1$ and $m\rightarrow m-2$ sideband transitions compared
to the $m\rightarrow m+1$ transition indicates
already a high population of the ground state of axial motion (along $y$) for the EIT-
cooled atoms. Performing robust adiabatic passages on the $m\rightarrow m-1$
transition, we infer, for the EIT-(molasses-)cooled atoms, a steady-state temperature of $(7.0\pm0.5)\,\mu$K ($(31\pm6)\,\mu$K),
corresponding to a final ground state occupation of $0.78\pm0.02$ ($0.29\pm0.05$), respectively. From the variation of the duration
of the EIT-cooling, a cooling rate of about $1\,$kHz has been inferred,
compatible with predictions in Ref.~\cite{Mor03}.

\section{Discussion}

With our setup we have successfully demonstrated
EIT-cooling applied to neutral atoms in a dipole trap. When
comparing the measurement of the survival probability with the
corresponding one in presence of the resonator, there is an
obvious difference on the red side ($\delta_\text{PL}<0$) of the EIT condition. This is because the cavity adds a new resonance, extending the cooling region to a larger range of probe-cavity detunings $\delta_\text{PC}$, thereby making the cooling more robust. For
this reason, these dynamics are also suitable for simultaneously
cooling several degrees of freedom, like for instance atomic
arrays~\cite{Lin13,SSm11}. Moreover, as shown in
Fig.~\ref{narrow_range}(b), in CEIT different cooling profiles can
be achieved by changing $\delta_{\rm LC}$, reflecting the
additional interference effect and resonance which are absent in
free space.

In conclusion, this work investigates the new scheme of cavity
assisted EIT cooling and provides a comparison between
the efficiency of free-space cooling and cavity cooling. These
techniques can be extended for cooling optomechanical
systems~\cite{*[{see, for example }] [{ and references therein; }]
kippenberg:2008,*chan:2011} coupled to a single
emitter~\cite{breyer:2012}. Our setup, moreover, can serve as a
transducer between the vibrational, electronic and photonic
degrees of freedom, thereby realizing a continuous-variable
quantum interface with single atoms~\cite{Parkins:1999}.

\section{Acknowledgments}
The authors acknowledge discussions with J. Eschner and D.
Leibfried, and financial support from BMBF (QuOReP), the European
Commission (IP SIQS, ITN CCQED, STREP PICC); and the German Research
Foundation (DFG). RR and SM acknowledge financial support from the
Studienstiftung des deutschen Volkes. MD, RR and SM acknowledge
financial support from the BCGS.


%

\end{document}